# Analyzing Black Hole super-radiance Emission of Particles/Energy from a Black Hole as a Gedankenexperiment to get bounds on the mass of a Graviton

A. Beckwith[1]

1) abeckwith@uh.edu, Chongqing University department of physics; Chongqing, PRC, 400044;

**Abstract**
Use of super-radiance in BH physics, so dE/dt < 0 specifies conditions for a mass of a graviton being less than or equal to 10^ - 65 grams, and also allows for determing what role additional dimensions may play in removing the datum that massive gravitons lead to 3/4[th] the bending of light past the planet Mercury.The present document makes a given differentiation between super-radiance in the case of conventional BHs and Braneworld BH super-radiance, which may delineate if Braneworlds contribute to an admissible massive graviton in terms of removing the usual problem of the 3/4[th] the bending of light past the planet Mercury which is normally associated with massive gravitons. This leads to a fork in the road, between two alternatives with the possibility of needing a multiverse containment of BH structure, or embracing what Hawkings wrote up recently, namely a re do of the Event Horizon hypothesis as we know it.

**Key words:** *massive gravity, Multiverse embedding structure* , re-do of Event Horizon

## 1. Introduction: Massive gravity and how to get it commensurate with black hole physics?

We are now attemping to come up with a critieria for either massless or massive gravitons. Our preferred way to do it is to distinguish between two forms of super-radiance. One built about Kerr black holes[1] , and the other involving brane theory [2], with the brane theory version of super radiance perhaps correcting a problem as to when a massive graviton would without brane theory, lead to 3/4th the angular bending of light, seen expermentally. We briefly allude to both of these cases in the introduction below, before giving more details as to this phenomenon in Section 2 and Section 3 below.

In general relativity the metric $g_{ab}(x,t)$ is a set of numbers associated with each point which gives the distance. to neighboring points. I.e. general relativity is a classical theory. As is designated by GR traditionalists [3], the graviton is usually stated to be massless. With spin two and with two polarizations. Adding a mass to the graviton results in 5 polarizations plus other problems [4,5] i.e. in [4], there is a description of how a massive graviton leads to 3/4th the calculated bending of light pass the mass of Mercury, as seen in the 1919 experiment . Reference [5] has details as to the five polarization states, which are another problem. One cannot go from a massive graviton, and eliminate mass from the graviton and then neatly recover the easier spin dynamics (2 polarization states) and vastly simpler situaiton where one has recovered the Schwartzshield metric. As reference [5] discusses, in its page 92, this easy recovery of the Schwartzshield metric if a graviton mass goes to zero, is impossible. Also note I.e. Note [4] has a discussion as to how the bending of light is not commensurate with GR for massive graviton, which is equivalent to a discussion as to a phenomenological ghost state for the trace of h, is given by [6] occurs regardless of if the mass for graviton nearly goes to zero . In reference [7] Czaki, Erlich, and Hollowood have given a temporary fix as to restoring the bending of light for massive gravitons to remove the 3/4th angle deflection from the 1919 GR test value, and this by use of brane theory.What this document will do will be to try to establish massive gravitons as super – radiant emission candidates from black holes [8] and in doing so provide another framework for their analysis which would embed them in GR. In doing so, one should keep in mind that this is a thought experiment and that the author is fully aware of how hard it would be to perform experimental measurements. In coming up with criteria as to graviton mass, we are also, by extension considering the Myers-Perry higher dimensional model of black holes [9] and commenting upon its applications, some of which are in [10] All of which start with the implications of dE/dt < 0, leading to 'leakage' from a black hole. i.e. energy of the black hole 'decreases' in time. I.e. there are ghost states, where h is the trace of h(i,j) which is a GW perturbation of the flat Euclidian metric, a possibility that brane theory and higher dimensions may remove the 3/4[th] angle of bent light calculated for massive gravitons, and a suitable thought experiment as given below may allow for dE/dt allowing us to determine if higher dimensional models are justifiable. This is the reason for the super-radiance phenomenology being investigated. i.e. of bending of angle of light divergence from GR models using massive Gravitons. Does dE/dt < 0 imply that there are brane theory states which may remove the 3/4th bending of light divergence from GR by massive gravitons? And can a super-radiance model for when dE/dt < 0 imply conditions for which brane worlds have to be considered [2], as opposed to the simpler model proposed by Padmanabhan [1]

The paper will differentiate between Kerr BH [1] versions of super-radiance and Brane theory BH super-radiance.[2] and secondly, aftewards inquire about if a BH in Brane theory configuration is satisfied if the simper Kerr BH super-radiance criteria is not satisfied.After these two versions are distinguished, we will then discuss experimental criteria which may result in determining if Kerr BH super radiance occurs [1], or Brane theory super radiance occurs[2], If only Kerr BH super radiance occurs, the likelyhood of massive Gravitons is remote. If brane theory BH super – radiance occurs, then there may[2] be conditions for which the 3/4th error in light bending is removed, permitting massive Gravitons.

## 2. What is super-radiance in black hole physics?  First the Padmanabhan treatment, for Kerr BHs.

We first of all consider a simplified version of super-radiance. In simple language, super-radiance involves having incoming radiation scattered off the horizon of a BH, and radiated outward, so the net flow of energy is dE/dt < 0 for 'radiation energy with a frequency bounded by $0 < \omega < m \cdot \Omega_H$ [1] . In this case m is a quantum number, the frequency $\omega$ is for radiation infalling to the event horizon of the BH, and the term, $\Omega_H$ is the angular velocity of a KERR black hole [2] . This paper first of all examines Padmanabhan's derivation of super-radiance [2] stating its application to the graviton, with mass, and making then a referral to the likelihood of measurement which ties in with the metric $g_{\mu\nu}$ being perturbed from flat space values by $h_{00}$, $h_{0i}$, and $h_{ij}$ [7], thereby making the case, due to the mass dependence of the black hole, that super-radiance would almost certainly not be observable but would firmly embed massive gravitons in GR in spite of the view point offered in [3]. To do so would mean that [1] has: if **dE/dt** < 0 for super-radiance, i.e. scattering of matter/energy from the horizon of a  BH, we examine, if $c_1$ is a constant, and $\omega$ a radiation frequency, and $m$ a quantum number and $\Omega_H$ angular velocity of the black hole. Here, after the Padmanabhan derivation of what dE/dt < 0 means, there will be a separate, brane theory derivation of BH super radiance [2] which has, provisionally $0 < \omega < \sum_{J=1}^{N/2} m_J \cdot (\Omega_H)_J$ [2], where N is tne number of dimensions. The 2$^{nd}$ frequency dependence for when N can go up to at least 10, or so will be remarked after we finish the Padmanbhan frequency dependence for super-radiance, as given below for a 'classical' Kerr BH. To start it off with, .look at [1]

$$\frac{dE}{dt} = c_1 \cdot \omega \cdot [\omega - m \cdot \Omega_H] \qquad (1)$$

In this case, according to Padmanabhan , $c_1$ is a constant, which is defined via writing Eq. (1) via[1]

$$\frac{dE}{dt} = \frac{M \cdot r_H}{2\pi} \cdot \left[\int \left(S^2(\theta) \cdot \sin^2\theta \cdot d\theta\right) \cdot d\phi\right] \cdot \omega \cdot [\omega - m \cdot \Omega_H]$$
$$= flux - of - energy - through - horizon \qquad (2)$$
$$\Leftrightarrow c_1 \equiv \frac{M \cdot r_H}{2\pi} \cdot \left[\int \left(S^2(\theta) \cdot \sin^2\theta \cdot d\theta\right) \cdot d\phi\right] = const.$$

Where for a massless scalar field one has the function S for a 'surface area'function, defined as follows, [1]

$$\exists S^2(\theta) \Leftrightarrow (-g)^{-1/2} \partial_b \left[ (-g)^{-1/2} \cdot g^{ab} \cdot \partial_a \Phi \right] = 0 \Leftrightarrow \Phi \equiv e^{-i\omega t} e^{im\cdot\phi} \cdot R(r) \cdot S(\theta) \quad (3)$$

In this case, mass M is for the source, i.e later for the mass M of a GW generator, in this case a BH. Also, here

$r_H$ is the horizon radius. As specified. And this will have its application to the issue of gravitons of a small mass spiriaing into a BH, with the BH subsequently releasing radiation via dE/dt < 0 as follows, with the given versions of a BH set of parameters [1] for a KERR BH. The following comes as far as angular velocity of the BH, as well as the following sets of parameters. Here, the phenomenon of super-radiance is impossible, if Eq.(6) below is zero. More on this later[1].

$$\Omega_H = \frac{a}{2M_{BH} \cdot r_H} \quad (4)$$

$$r_H = M_{BH} - \sqrt{M_{BH}^2 - a^2} \quad (5)$$

$$a = \sqrt{x^2 + y^2} \quad (6)$$

Then,

$$0 < \frac{\omega}{m} < \Omega_H \quad (7)$$

Note that there are conditins, based upon Eq. (4) above, which go to zero, due to the numerator, in a manner which means that BH physics, for 'simpler' BH physics one can have for the simplest situation one where there is no angular rotation of the event horizon, and thereby no super-radiance. This will obviously lead to the classical description of BH physics. Problem though, is that recently, Hawkings has stated that not all is well in BH event horizons, and that scrambled information could possibly leave a BH, in opposition

## 2a. Examining Super-radiance when there is more than 4 dimensions as to BH physics.

As said before, [2]

$$0 < \omega < \sum_{J=1}^{N/2} m_J \cdot (\Omega_H)_J \quad (8)$$

In doing so, N as given above is a measure of dimensions as to the BH , and the difference in this from Eq. (7) in part is also due to

$$\Omega_H = \left. \frac{a}{2M_{BH} \cdot r_H} \right|_{Kerr-BH} \xrightarrow{Kerr-BH \to Myers-Perry-BH} \left. \frac{a}{a^2 + r_H^2} \right|_{Myers-Perry-BH} \quad (9)$$

The numerator of the above is still defined by the square root of x^2+y^2, and could go to zero for certain quantum mumbers, $m_J$, and we would then paraphrase the right hand side of Eq. (9) as functionally being

$$\Omega_j = \frac{a_J}{a_J^2 + r_H^2} \text{ as } \textit{frequency of BH arising due to the jth component of BH angular Momentum } J_j \quad (10)$$

So then one has a re write of Eq. (8) as given, with a slightly different angular frequency for BH's as by [2]

$$0 < \omega < \sum_{j=1}^{N/2} m_j \frac{a_j}{a_j^2 + r_H^2}\bigg|_{Myers-Perry-BH} . \tag{11}$$

This is to be compared with the Padmanabhan version of super-radiance as given by:[1]

$$0 < \omega < m \cdot \frac{a}{2M_{BH} \cdot r_H}\bigg|_{Kerr-BH} \tag{12}$$

We will be commenting upon what the experimental signatures of both Eq. (11) and Eq. (12) could be, and why. In the next section.

## 3. Could super radiance be observed experimentally, and what good is this thought experiment?

Super radiance is really about the same as particle production from a BH. From Padmanabhan [1], page 623, comes the result, that as given in a quote, after [1]s page 623 quote, as follows: namely

***If we think of super radiance as stimulated emission of radiation by the black hole in certain modes, owing the presnce of the incoming wave, it seems natural to expect spontaneous emission of radiation in various modes by the black hole in quantum field theory. The black hole evaporation (then) can be thought of as spontaneous emission of particles that survives even in the limit of zero angular momentum of the black hole.***

Furthermore, on the same page, page 623 of [1] has that

***It seems natural to assume that this source of energy radiated to infinity is the mass of the collapsing structure.***

Leading to, Formula 14.143 of reference [1] that the 'entropy' of a BH is given by, where M is the mass of the BH, $L_P$ Plamck length, and $A_{hor}$ is the area of the Event Horizon of a black hole. This area of a BH event horizon is relevant since it directly connects, as we will mention later, to the reference [2] version of super-radiance. Reference [1] 's version of entropy would also hold for [2] as well, and we state the entropy as

$$S = 4\pi M^2 = \frac{1}{4} \cdot \left(\frac{A_{hor}}{L_P^2}\right) \tag{13}$$

Here, in reference [2] we have that in its ( reference [2]) equation 24, that its main result is about the differential of the area of an event Horizon which is given as, if there is a Brane theory connection to the formation of BHs

$$dA_{hor} = \frac{8\pi r_H}{B} dM_{BH} \cdot \left(1 - \frac{1}{\omega} \sum_{j=1}^{N/2} m_j \cdot \Omega_j\right) \tag{14}$$

The postive definite nature of this expression for the differential of the area of an event Horizon would then be [2] since, dM < 0, then by [2], so then by Eq.(15) below, we recover Eq.(11), by [2], if dA >0, then

$$dM_{BH} \cdot \left(1 - \frac{1}{\omega}\sum_{j=1}^{N/2} m_j \cdot \Omega_j\right) > 0 \Leftrightarrow \left(1 - \frac{1}{\omega}\sum_{j=1}^{N/2} m_j \cdot \Omega_j\right) < 0 \tag{15}$$

We make the following 3 claims as for the analogy to BH physics, namely:

Claim I . Entrophy in both Kerr and Myers-Perry BHs have dA >0 , where A is the event horizon, and

  I.  For Myers-Perry BHs, the following are true (dimensions up to 10, say, i.e. N = 10).

$$S = 4\pi M^2 = \frac{1}{4} \cdot \left( \frac{A_{hor}}{L_P^2} \right)$$

$$dA_{hor} = \frac{8\pi r_H}{B} dM_{BH} \cdot \left( 1 - \frac{1}{\omega} \sum_{j=1}^{N/2} m_j \cdot \Omega_j \right)$$

$$\left( 1 - \frac{1}{\omega} \sum_{j=1}^{N/2} m_j \cdot \Omega_j \right) < 0$$

  II.  For Kerr BH one could arguably have much the same thing. I.e.

$$S = 4\pi M^2 = \frac{1}{4} \cdot \left( \frac{A_{hor}}{L_P^2} \right)$$

$$dA_{hor} = \frac{8\pi r_H}{B} dM_{BH} \cdot \left( 1 - \frac{1}{\omega} m \cdot \Omega_H \right)$$

$$\left( 1 - \frac{1}{\omega} m \cdot \Omega_H \right) < 0$$

*Proof : Eq. (15) as well as applying Eq. (9), Eq.(10),Eq.(11).*

*Claim 2, If a is zero, then super radiance as made possible inClaim 1, part II is impossible for Kerr Black holes*

*Proof: a goes to zero, mean numerator of Eq. (9) goes to zero. Hence,* $\left( 1 - \frac{1}{\omega} m \cdot \Omega_H \right) < 0$ *does not happen. Hence, for non zero frequency of incoming radiation,* $0 < \omega < m \cdot \Omega_H$ *does not hold. Hence no BH super-radiance.*

*Claim 3. One could have the following: Claim 1, part II, may be false, but Claim 1, part I may be true*

*Proof: For* $N \geq 4$ *or so, the following decomposition may be true,*

$$\left( 1 - \frac{1}{\omega} \sum_{j=1}^{N/2} m_j \cdot \Omega_j \right) = \left[ 1 - \frac{1}{\omega} m \cdot \Omega_H \right] - \frac{1}{\omega} \sum_{j=2}^{N/2} m_j \cdot \Omega_j < 0$$

If the first term in [ ] in the RHS of the above formula is = 0, claim I, part II is false, but one could still have Claim I, part I as true. i.e. one could write the following.

$$0 < \omega < \sum_{J=2}^{N/2} m_J \cdot \left( \Omega_H \right)_J$$ Then claim I, part I will be true, i.e. super-radiance for Brane theory BHs

The significance of the three claims is as follows. As given by reference [4], there is a problem, if a massive graviton exists, the bending of light, say about Mercury, the Eddinton 1919 experiment is calculated to be 3/4[th] the value seen in the 1919 experiment which proved classical GR. By reference [7] there can be a situation for which if there exists higher than 4 dimensional brane theory, one may correct the 3/4[th] deficiency. But if Claim 1, part I is not true, then the solution allowing for [7] likely is not true.

Note that the super-radiance phenomenon as referenced in Claim I, part 1 and part 2, has its roots in ENTROPY. Note that entropy of a black hole with its surface area is stated to be a pre condition for initial conditions for super-radiance. And more than that, one needs a spinning black hole. No black hole spin, with a commensurate treatment could lead to just black hole evaporation, as noted above, but BH evaporation is not the same as the super – radiance phenomenon.

## 3a. Minimum experimental bounds which may effect the results of our inquiry, provided Claim I, part I is true, and perhaps, Claim 3 true also holds. I.e. Myers-Perry as a black hole higher dimensional representation which may permit massive gravitons.

IMO, as stated above, the Meyers- Perry condition for BHs is, as a gateway a probable candidate as to experimental observations for BHs. As mentioned earlier, for higher dimensional BHs which may allow for massive gravitons, here are the perturbations due to GW due to higher dimensional black holes. We state these as follows, namely

The subsequent values by $h_{00}$, $h_{0i}$, and $h_{ij}$ make the case, due to mass dependence of the black holes in the Myers-Perry black holes, be a challenge. [4] has

$$h_{00} \approx \frac{16\pi G}{(d-2)\cdot \Omega_{d-2}} \cdot \frac{M_{BH}}{r^{d-3}}$$
$$h_{ij} \approx \frac{16\pi G}{(d-2)\cdot (d-3)\cdot \Omega_{d-2}} \cdot \frac{M_{BH}}{r^{d-3}} \cdot \delta_{ij} \qquad (16)$$
$$h_{oi} \approx -\frac{8\pi G}{\Omega_{d-2}} \cdot \frac{x^k}{r^{d-1}} \cdot J^{ki}$$

The coefficient d is for dimensions, 4 or above, and in this situation, with angular momentum $J^{ki}$. Here the term put in, namely Eq/ (17) is for angular area, and it has NO relationship to the formula for angular velocity of BHs, namely Eq. (17) has NO RELATIONS to Eq. (4) and Eq. (9) above.

$$\Omega_{d-2} = 2\pi^{(d-1)/2} \Big/ \Gamma\left(\frac{d-1}{2}\right) \qquad (17)$$

$$J^{ki} = 2\cdot \int x^k \cdot T^{i0} \cdot d^{d-1}x \qquad (18)$$

The $T^{i0}$ above is a stress energy tensor as part of a d dimensional Einstein equation given in [5] as

$$R_{jl} - \frac{1}{2}\cdot g_{jl}\cdot R = 8\pi G T^{jl} \qquad (19)$$

Also, the mass of the black hole is, in this situation scaled as follows: if $\mu$ is a re scaled mass term[5]

$$M_{BH} = \mu \cdot \Omega_{d-2}\cdot (d-2)/16\pi G \qquad (20)$$

More generally, the mass of the black hole is written as

$$M_{BH} \equiv \int T_{00} d^{d-1}x \qquad (21)$$

We will next go to the minimum size of a black hole which would survive as up to 13.6 billion years, and then say something about the relative magnitude of the magnitude of the terms in Eq.(16) and then their survival today, and what that portends as to the strength of signals which may be received. The variance of black hole masses, from super massive BHs to those smaller than $10^{15}$ *grams* will be discussed, in the context of Eq.(16), and stress strength, with commentary as to what we referred to earlier, namely strain for detecting GW is given by h(t) given below, with $D^{ij}$ as the detector tensor, i.e. a constant term, so that by [4], page 336, we write

$$h(t) = D^{ij} h_{ij} \qquad (22)$$

This Eq.(22) means that the magnitude of strain, h, is effected by Eq. (16) ,Eq.(17) and Eq. (18). and its magnitude, seen next. Note that the magnitude of the strain, h, as being brought up may be affected by the mass of a graviton, due to T, which is a feed into Eq. (16) above. Nalmely consider that the mass assumed for the graviton is of the order of 10^-65 grams, which is given by [5] by, if h does not equal zero, then the stress energy tensor of the massive graviton is for non zero $T_{uv}$ which corresponds to a non zero concentration in interstellar space, with[5]

$$m_g^2 = -\frac{\kappa}{6h} T$$
$$T = traceT_{uv} \qquad (23)$$

We will get explicit upper bounds to Eq. (23), and use it as commentary in the conclusion of this article. That will affect the infalling frequency $\omega$ which will be part of the super-radiance discusion.

## 3b. Values of the Meyers- Perry $h_{00}$, $h_{0i}$, and $h_{ij}$ in magnitude lead to nominal h values.

If *D* below is redshift corrected distance, in a rough sense, leads to an approximation of h as roughly proportional to $h_{00}$ with the roughly scaled results of

$$h \sim GM/c^2 D \qquad (24)$$

Note that if the tensor $D^{ij}$ is approximately unity, with then the results as given by

$$M_{BH}\big|_{min-life.time} \propto 10^{15} \, grams \Leftrightarrow h_{00} \, \& \, h_{ii} \propto 10^{-40} \text{ for BHs ; Z(redshift)~10} \qquad (25)$$

Whereas super massive black holes, of about 100 times solar mass at Z(redshift)~10

$$M_{BH}\big|_{100-solar-mass} \Leftrightarrow h_{00} \, \& \, h_{ii} \propto 10^{-20} \qquad (26)$$

It is easy from inspection to infer from this that most early formed black holes would not be accessible and that only the giant ones would do. With that, we next then explore the frequency ranges which could lead to certain Graviton masses, as could be linked to super-radiance. I.e. it would mean a very large SMBH, of about 100 solar masses of a redshift of the order of Z ~ 10 at or less than a billion years after the creation of the universe, would lead to the values of Eq. (26) above, which could be conceivably detected, which then leads us to the question of what frequencies of the graviton, if presumably massive would be involved. This would then allow us to make inquiry as to what the Meyers-Perry values for super-radiance and absorption/subsequent reflection of GW radiation which could conceivably be detected for strain values of the order given by Eq. (26) above.

## 3c. Frequency and wavelengths for ultra low 'massive graviton' masses.

To get the appropriate estimates, we turn to reference [11], by Goldhaber, and Nieto, which can be used to give a set of frequency and mass equivalences for the 'massive' graviton, on the order of having the following equivalent values as paired together, namely starting off, with graviton mass, graviton wavelength, and resulting graviton frequency, we observe the dual pairing of: if one also looks at Valev's estimates also, to get [12]

$$m_g \sim 2 \times 10^{-65} \, grams$$
$$\lambda_g \sim 2 \times 10^{22} \, meters$$
$$\sim 10^{-4} \cdot radius-of-universe \quad (27)$$
$$\omega_g \sim (3/2) \times 10^{-14} \, / \, second$$

Obviously, with regards to this, if such an extremely low value for resultant frequency is obtained, and then one is obtaining, the value that it is inevitable, just in terms of frequency, to have for any spinning Kerr BH

$$\omega_g \sim (3/2) \times 10^{-14} \, / \, second < \Omega_H \quad (28)$$

If we treat, then for any reasonable non zero value of $a$ we will find that for a SMBH of about 100 Solar masses, one will still have, realistically, $\omega_g \sim (3/2) \times 10^{-14} \, / \, second \ll \Omega_H$. The author has found though, that for SMBHs of say a million or more times the mass of the sun, that in say the spiral galaxy SMBH in the center that instead of an easy upper bnd, $\omega_g \sim (3/2) \times 10^{-14} \, / \, second \leq \Omega_H$ which leads to Claim 4,

**Claim 4, for Super-radiance ( Kerr style)**

$\omega_g \sim (3/2) \times 10^{-14} \, / \, second \ll \Omega_H$ (easy Super-radiance) for SMBH 100 times solar mass

$\omega_g \sim (3/2) \times 10^{-14} \, / \, second \leq \Omega_H$ (problematic Super –radiance) for SMBH 10 ^ 6 times solar mass.

The **proof** is in the definition of $\Omega_H = \left. \frac{a}{2M_{BH} \cdot r_H} \right|_{Kerr-BH}$. With a very small numerator.

This claim 4, means that before the formation of massive spiral galaxies, that super – radiance is doable. However, the author fails to understand how it is possible on another theoretical ground, i.e. what does super raidance mean for BHs for which

$$\lambda_g \sim 2 \times 10^{22} \, meters$$
$$\sim 10^{-4} \cdot radius-of-universe \quad (29)$$

On the face of it, this is absurd. I.e. how could wavelengths the 1/10,000 the size of the universe interact with a Kerr Black hole ?

We claim, that the embedding of black holes in five or higher dimensional dimensional space time is a way to make a connection with a multiverse, as given in the following supposition [13]/ and that this may be the only way to reconcile what seems to be an absurd proposition. I.e. graviton wavelengths 1/10,000 the size of the standard 4 dimensional universe interacting with spinning black holes in 4 dimensional space – time, whereas that moderate 100 times the mass of the sun BHs easily satisfy $\omega_g \sim (3/2) \times 10^{-14} / \text{second} \ll \Omega_H$.

**4. Conclusion, in one way rediculously easy to obtain super-radiance for massive gravitons, and another sense an absurd proposition. Could a multiverse embedding of BHs be a way out of what otherwise seems an impossible Dichotomy ? Or will we have to embrace Hawking's suggestion that the event Horizon has foundationally crippling flaws?**

To address this problem, the author looks at two suggestions. Either the BH are really embedded in a multiverse, and has a different geometry in higher dimensions than is supposed, or one goes to the recent Hawkings hypothesis which changes entirely the supposition of the Event horizon. Namely

We will first of all give a brief introduction to the Penrose CCC hypothesis generalized to a multiverse.

**4a. Extending Penrose's suggestion of cyclic universes, black hole evaporation, and the embedding structure our universe is contained within, i.e. using the implications of Eq.(29) for a multi verse. This multiverse embedds BHs and may resolve what appears to be an impossible dichotomy.**

That there are no fewer than N universes undergoing Penrose 'infinite expansion' (Penrose, 2006) [13] contained in a mega universe structure. Furthermore, each of the N universes has black hole evaporation, with the Hawking radiation from decaying black holes. If each of the N universes is defined by a partition function, called $\{\Xi_i\}_{i=N}^{i=1}$, then there exist an information ensemble of mixed minimum information correlated as about $10^7 - 10^8$ bits of information per partition function in the set $\{\Xi_i\}_{i=N}^{i=1}\Big|_{before}$, so minimum information is conserved between a set of partition functions per universe

$$\{\Xi_i\}_{i=N}^{i=1}\Big|_{before} \equiv \{\Xi_i\}_{i=N}^{i=1}\Big|_{after} \tag{30}$$

However, there is non-uniqueness of information put into each partition function $\{\Xi_i\}_{i=N}^{i=1}$. Furthermore Hawking radiation from the black holes is collated via a strange attractor collection in the mega universe structure to form a new big bang for each of the N universes represented by $\{\Xi_i\}_{i=N}^{i=1}$. Verification of this mega structure compression and expansion of information with a non-uniqueness of information placed in each of the N universes favors ergodic mixing treatments of initial values for each of N universes expanding from a singularity beginning. The $n_f$ value, will be using (Ng, 2008) $S_{entropy} \sim n_f$. [14] . How to tie in this energy expression, as in Eq. (30) will be to look at the formation of a nontrivial gravitational measure as a new big bang for each of the N universes as by $n(E_i)$. the density of states at a given energy $E_i$ for a partition function. (Poplawski, **2011**) **[15]**

$$\{\Xi_i\}_{i=1}^{i=N} \propto \left\{\int_0^\infty dE_i \cdot n(E_i) \cdot e^{-E_i}\right\}_{i=1}^{i=N}. \tag{31}$$

Each of $E_i$ identified with Eq.(31) above, are with the iteration for N universes (Penrose, 2006)[14] Then the following holds, namely

**Claim 5,**

$$\frac{1}{N} \cdot \sum_{j=1}^{N} \Xi_j \Big|_{j-before-nucleation-regime} \xrightarrow{vacuum-nucleation-tranfer} \Xi_i \Big|_{i-fixed-after-nucleation-regime} \quad (32)$$

For N number of universes, with each $\Xi_j \Big|_{j-before-nucleation-regime}$ for j = 1 to N being the partition function of each universe just before the blend into the RHS of Eq. (32) above for our present universe. Also, each of the independent universes given by $\Xi_j \Big|_{j-before-nucleation-regime}$ are constructed by the absorption of one to ten million black holes taking in energy. **I.e. (Penrose, 2006)** [14]. Furthermore, the main point is similar to what was done in [18] in terms of general ergodic mixing

**Claim 6**

$$\Xi_j \Big|_{j-before-nucleation-regime} \approx \sum_{k=1}^{Max} \tilde{\tilde{\Xi}}_k \Big|_{black-holes-jth-universe} \quad (33)$$

What is done in **Claim 5 and Claim 6** is to come up with a protocol as to how a multi dimensional representation of black hole physics enables continual mixing of spacetime [16] largely as a way to avoid the Anthropic principle, as to a preferred set of initial conditions. With investigations, this complex multiverse may allow bridging what seems to be an unworkable dichotomy between ultra low graviton frequency, corresonding roughly to 10^-65 grams in rest mass , easily satisfied by Kerr black holes with rotational frequencies, as given in out text as many times greater, combined with the absurdity of what is Eq. (29). How can a graviton with a wavelength 10^ - 4 the size of the universe interact with a Kere black hole, spatially. Embedding the BH in a multiverse setting may be the only way out.

**Claim 5** is particularly important. The idea here is to use what is known as CCC cosmology, which can be thought of as the following.

First. Have a big bang ( initial expansion) for the universe. After redshift z = 10, a billion years ago, SMBH formation starts. Matter- energy is vacuumed up by the SMBHs, which at a much later date than today ( present era) gather up all the matter-energy of the universe and recycles it in a cyclic conformal translation, as follows, namely

$$E = 8\pi \cdot T + \Lambda \cdot g$$
$$E = source\ for\ gravitational\ field$$
$$T = mass\ energy\ density \quad (34)$$
$$g = gravitational\ metric$$

$$\Lambda = vacuum\ energy, rescaled\ as\ follows$$

$$\Lambda = c_1 \cdot [Temp]^\beta \quad (35)$$

C1 is , here a constant. Then

The main methodology in the Penrose proposal has been in Eq. (35) evaluating a change in the metric $g_{ab}$ by a conformal mapping $\hat{\Omega}$ to

$$\hat{g}_{ab} = \hat{\Omega}^2 g_{ab} \tag{36}$$

Penrose's suggestion has been to utilize the following[17]

$$\hat{\Omega} \xrightarrow{ccc} \hat{\Omega}^{-1} \tag{37}$$

Infall into cosmic black hopes has been the main mechanism which the author asserts would be useful for the recycling apparent in Eq.(37) above with the caveat that $\hbar$ is kept constant from cycle to cycle as represented by

$$\hbar_{old-cosmology-cycle} = \hbar_{present-cosmology-cycle} \tag{38}$$

Eq. (37) is to be generalized, as given by a weighing averaging as given by Eq.(32) where the averaging is collated over perhaps thousands of universes, call that number N, with an ergotic mixing of all these universes, with the ergotic mixing represented by Eq.(32) to generalize Eq.(37) from cycle to cycle.

4b/If this does not work, and the multivese suggestion is unworkable, there then has to be a consideration of the zero option. **Namely Hawkings throwing out the Event Horizon as we know it in BH physics**. See this reference, namely [18]

We are in for interesting times. I see turbulence and interesting results ahead

Acknowledgements

**This work is supported in part by National Nature Science Foundation of China grant No. 11375279**## Bibliography

**This work is supported in part by National Nature Science Foundation of China grant No. 11375279**

[1] T. Padmanabhan *Gravitation, Foundations and Frontiers* Cambridge University Press, NYC, NY, USA 2010

[2]E. Jung,S.Kim,D.K. Park, "Conditions for the Super-radiance Mode in Higer Dimensional Rotating Black Holes with Angular Momentum Parameters",Phys.Lett.B.619,(2005), pp. 345-351, Arxiv:hep-th/0504139

[3] S. Weinberg, *Gravitation and Cosmology: Principles and Applications of the General Theory of Relativity*: Wiley Scientific, 1972

**[4]** P. Tinyakov, "Giving mass to the Graviton", pp 471-499, as part of Les Houches, Session **LXXXVI**, *Particle Physics, and Cosmology, the fabric of Space-time*, F. Bernadeau, C. Grojean, and J. Dalibard, Eds, of Elsevier Press, Oxford, UK, **2007**

**[5]** M. Maggiore, *Gravitational Waves: Volume 1. Theory and Practice* Oxford University press, 2008, Oxford UK

[6] P. Creminelli, A. Nicholis, M. Papucci, and E. Trincherini (2005), JHEP 0509, 003

[7] C. Cesaki, J. Erlich, T. Hollowood,"Graviton Propagators, Brane Bending and Bending of Light in Theories with Quasi Localized Gravity", arXIV Hep-th/0003020 v1, Phys. Lett B. 481: pp 107-113, 2000

[8] M. Meyers *Meyers – Perry Black holes*, pp 101-131 in *Black holes in higher dimensions* edited by G. Horowitz, Cambridge University Press, NYC, NY, USA, 2012

[9] Entire book: *Black holes in higher dimensions* edited by G. Horowitz, Cambridge University Press, NYC, NY, USA,